\documentclass[a4paper,fleqn,usenatbib]{mnras}

\usepackage{newtxtext,newtxmath}
\usepackage[T1]{fontenc}
\usepackage{ae,aecompl}

\usepackage{graphicx}	% Including figure files
\usepackage{amsmath}	% Advanced maths commands
\usepackage{amssymb}	% Extra maths symbols

\title[The spatial substructure of Cygnus OB2]{Disentangling the spatial substructure of Cygnus OB2 from {\it Gaia} DR2}

\author[S.R.Berlanas]{
S. R. Berlanas,$^{1,2}$\thanks{E-mail: srberlan@iac.es}
N. J. Wright,$^{3}$
A. Herrero,$^{1,2}$
J. E. Drew$^{4}$ 
and D. J. Lennon$^{5,1}$ 
\\ \\
$^{1}$Instituto de Astrof\'isica de Canarias, 38200 La Laguna, Tenerife, Spain\\
$^{2}$Departamento de Astrof\'sica, Universidad de La Laguna, 38205 La Laguna, Tenerife, Spain\\
$^{3}$Astrophysics Group, Keele University, Keele, ST5 5BG, UK\\
$^{4}$School of Physics, Astronomy \& Mathematics, University of Hertfordshire, Hatfield AL10 9AB, UK\\
$^{5}$ESA, European Space Astronomy Centre, Apdo. de Correos 78, E-28691 Villanueva de la Ca\~nada, Madrid, Spain}
\date{Accepted XXX. Received YYY; in original form ZZZ}

\pubyear{2019}

\begin{document}
\label{firstpage}
\pagerange{\pageref{firstpage}--\pageref{lastpage}}
\maketitle

\begin{abstract}
For the first time, we have explored the spatial substructure of the Cygnus~OB2 association using parallaxes from the recent second {\it Gaia} data release. We find significant line-of-sight substructure within the association, which we quantify using a parameterised model that reproduces the observed parallax distribution. This inference approach is necessary due to the non-linearity of the parallax-distance transformation and the asymmetry of the resulting probability distribution. By using a Markov Chain Monte Carlo ensemble sampler and an unbinned maximum likelihood test we identify two different stellar groups superposed on the association. We find the main Cygnus~OB2 group at $\sim$1760 pc, further away than recent estimates have envisaged, and a foreground group at $\sim$1350 pc. We also calculate individual membership probabilities and identify outliers as possible non-members of the association.
%It should be a single paragraph not more than 250 words (200 words for Letters).
%No references should appear in the abstract.
\end{abstract}


%%%%%%%%%%%%%%%%%%%%%%%%%%%%%%%%%%%%%%%%%%%%%%%%%%

%%%%%%%%%%%%%%%%% BODY OF PAPER %%%%%%%%%%%%%%%%%%

\section{Introduction}

A key difficulty in the study of Milky Way massive stars and OB associations  has  been  the  large  uncertainty  in  their  distances,  hindering  the  comparison  with  theories  of  stellar  and  cluster  evolution. They are needed to place the stars in the Hertzsprung-Russell Diagram (HRD), obtaining a better comparison of stellar masses and radii derived from the spectroscopic analyses and the evolutionary codes (a persistent problem in the field of massive stars, see \cite{herrero92,repolust04,massey12,markova15}).

The  recent second data release (DR2) from the {\it Gaia}  satellite \citep[][]{prusti16, brown18} has provided unprecedented high quality astrometry for more than 1.3 billion objects, all with measured parallaxes. Parallax uncertainties (excluding a conservative systematic error up to 0.1 mas, see \cite{luri18}) are around 0.04 milliarcseconds (mas) for bright sources  (G $<$ 14 mag),  around  0.1  mas  for  sources with a G magnitude $\sim$ 17, and around 0.7 mas for the faintest (G $\sim$ 20 mag). This scenario provides a unique opportunity to inspect the internal structure of Galactic young open clusters and relatively nearby massive OB associations.

\begin{figure*}
\centering
	\includegraphics[width=12cm]{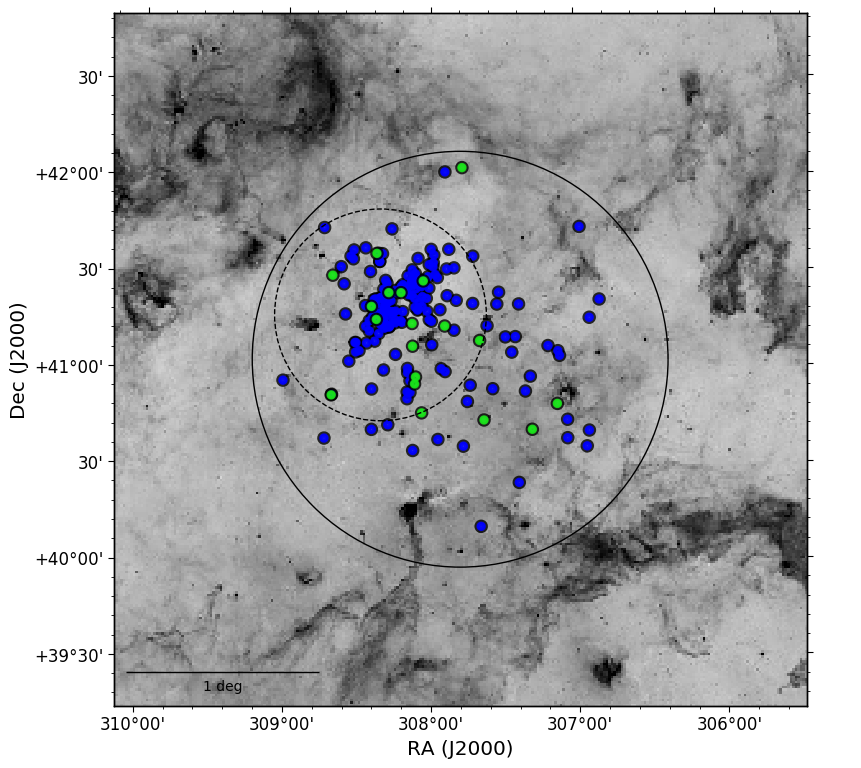}
    \caption{Inverse Spitzer 8 $\mu$m image showing the location of the two main stellar groups found in the region (see Sect.~\ref{sec:results} for further details). Blue colour represents stars from the main Cygnus~OB2 population and green those stars found to be in a foreground group. The solid line circle delimits the 1 degree radius area adopted in this work. For reference, the dash-dotted line circle shows the area considered by \citet{wright15} indicating the core of the association.}
    \label{fig:sample}
\end{figure*}

The Cygnus~OB2 association is one of the most massive OB associations at less than 2 kpc from the Sun \citep{knod03,rygl12}. Hosting  hundreds  of  OB  stars, it  is the  most  obvious  example  of  recent  star formation in the massive Cygnus-X complex. Its massive star population has been widely studied, including membership \citep{mt91, knod00, comeron02, hanson03, neg08,com12, berlanas18a}, mass function \citep{ kiminki07, wright15}, extinction \citep{ hanson03,guarcello12,com12,wright15} and chemical composition \citep{berlanas18b} studies. The distribution of stellar ages extends beyond 20 Myr \citep{com16} and a correlation between age and Galactic longitude exists, suggesting that massive-star formation has proceeded from lower to higher Galactic longitudes \citep{com12, berlanas18a}.  The significant spatial \citep{wright14} and kinematic substructure found by \cite{wright16} could indicate that Cygnus OB2 is made up of different individual subgroups. However, an uncertainty over whether all its OB stellar content are at the same distance persists. 
The high-precision {\it Gaia} DR2 parallaxes could therefore be used to properly study and unravel the spatial substructure of this association. Differentiating internal subgroups will help to understand the star formation process, origin, and evolution of the association, as well better characterize the stellar content in the region.

This paper is organized as follows. In Section~\ref{sec:data} we present the data and selection criteria. In Section~\ref{sec:method} the modelling approach used in this work is detailed. In Section~\ref{sec:results} we show the results of the best-fitting model and membership probabilities. A discussion of these results is provided in Section~\ref{sec:discussion}. Finally, we summarize the work in Section~\ref{sec:conclusions}.

\section{Data}\label{sec:data} 

\subsection{Stellar sample}\label{sec:selection} 

The sample of stars used for this study is comprised of known OB members of Cygnus~OB2 within a radius of 1 deg. of the coordinates $l = $79.8$^{\circ}$ and $b = $+0.8$^{\circ}$. We gathered stars from the samples of \citet{wright15} and \citet{berlanas18a}, the former of which is a census of spectroscopic members gathered from the literature \citep[e.g.,][]{mt91,comeron02,hanson03,kiminki07} while the latter expands this work to include more stars over a wider area. This produced a sample of 229 members of Cygnus~OB2, 167 of which are located in the core of the region (see  Fig.~\ref{fig:sample}).

\subsection{{\it Gaia} DR2 parallaxes}\label{sec:parallaxes}
\label{section_parallaxes}

Astrometry for this work was taken from {\it Gaia} DR2 \citep{brown18}.  We included stars that have astrometry that passed the selection criteria recommended by L. Lindegren based on the re-normalised unit weight error (or RUWE), defined as $u_{norm} = u / u_{0}(G,C)$ where $u = (\mathrm{\texttt{astrometric\_chi2\_al}} / \mathrm{\texttt{astrometic\_n\_good\_obs\_al}} - 5)^{1/2}$ and $u_{0}(G, C)$ is a smooth function in magnitude ($G$) and colour ($C = G_{BP} - G_{RP}$)\footnote{See technical note GAIA-C3-TN-LU-LL-124-01 available at https://www.cosmos.esa.int/web/gaia/public-dpac-documents}. We adopted RUWE $\leq$ 1.4 as the selection criterion for good astrometric solutions, as recommended in the above cited technical note. 
This cut caused us to discard 29 stars, resulting in a sample of 200 targets with reliable {\it Gaia} astrometry. We also note that all the targets of our sample meet with the \texttt{visibility\_periods\_used} $>$ 8 criterion, which is a key recommendation from the data release papers \citep{lindegren18, arenou18}.
The final stellar sample used for this work and those stars discarded by the selection criteria are available in electronic form at the CDS and at MNRASL online.

\begin{figure*}
 \centering
	\includegraphics[width=5.8cm,height=4.3cm]{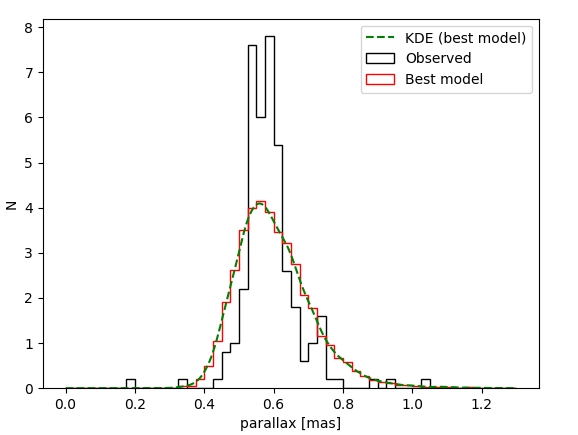}
	\includegraphics[width=5.8cm,height=4.3cm]{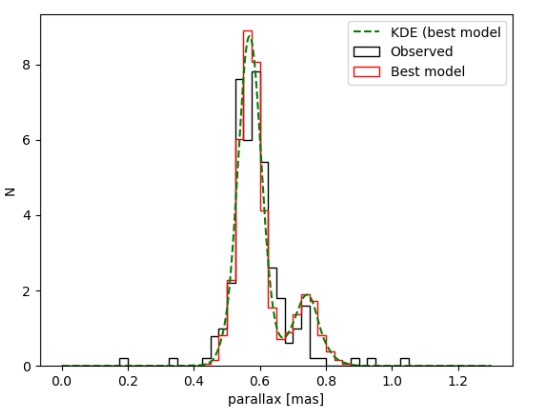}
    \includegraphics[width=5.8cm,height=4.3cm]{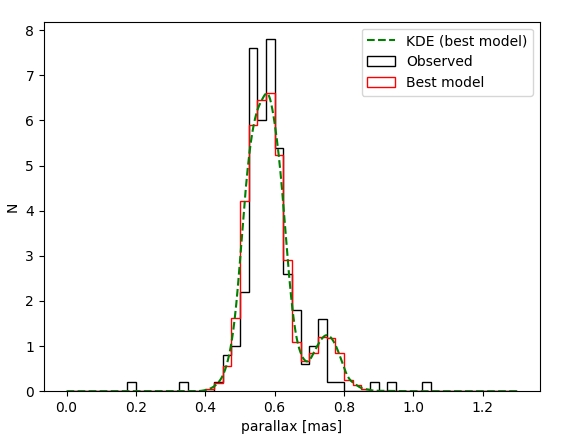}
    \caption{Normalized parallax distribution of the Cygnus~OB2 sources (in black) and the derived best-fitting models (in red). Green represents a kernel-density estimation using Gaussian kernels. Left, middle and right hand panels show the 1-, 2- and 3- component distributions, respectively.}
    \label{fig:models}
\end{figure*}

{\it Gaia} DR2 parallax uncertainties are derived from the formal errors computed in the astrometric processing. Additional systematic uncertainties of up to 0.1~mas exist and depend on factors such as the position on the sky, magnitude, and colour of the targets \citep{lindegren18}. Since our  goal is  not  to  obtain absolute distances for individual sources but to resolve internal substructure of the association we only consider the relative parallaxes of sources in the association. We do not expect the systematic error to vary across our sample since our field of view is relatively small (1 degree) and our sample has similar magnitudes and colours. Therefore, systematic parallax uncertainties are not included in our analysis, but are added when absolute distances are calculated \citep[as will the parallax zero point offset of -0.03~mas,][]{lindegren18}.

\section{Modelling method}\label{sec:method}

The observed parallax distribution of our sample (see Fig.~\ref{fig:models}, in black) peaks at about 0.6~mas, but is wider than would be expected if it's width was entirely due to parallax uncertainties. The distribution also shows evidence for multiple groups along the line-of-sight. Therefore, instead of estimating the distance to the association based on the average parallax we model the parallax distribution as a series of groups, each with an inherent width and different distance.

To infer the distance to the Cygnus~OB2 association we use a parameterised model of the distance to the association to reproduce the observed parallax distribution of the massive stars. The model predicts a distribution of parallaxes that is then compared to the observed distribution in parallax space. This Bayesian inference process is critical when using parallaxes because of the non-linearity of the transformation between these quantities and the asymmetry of the resulting probability distribution \citep{bailer15}.

We model the stellar population assuming it is composed of $N$ components, each of which contains a fraction of the total stellar content, $f_N$, and have distances that follow a Gaussian distribution. Each component therefore has free parameters for the centre, $d_N$, and standard deviation, $\sigma_N$, of each Gaussian, as well as an additional $N-1$ parameters to represent the fraction of stars in each component. Thus the model has a total of $3 N - 1$ parameters. We use wide and linear priors, allowing the central distances for each component of the association to vary in the range of 1--2 kpc and the standard deviations to vary from 0--1 kpc.

\begin{table}
\centering
\caption{Statistical data of the obtained Gaussian distributions based on 1-, 2- and 3- component best-fitting models.}
\label{table1}
\begin{tabular}{c c c c c c c }   
\hline\\[-1.8ex]
\small{} & \small{Model 1}& \multicolumn{2}{c}{\small{Model 2}}&  \multicolumn{3}{c}{\small{Model 3}}\\
\cline{2-7}\\[-1.5ex]
\small{$N$}  &\small{1} &\small{1}& \small{2}& \small{1}& \small{2} & \small{3}  \\    
\hline\\[-1.5ex]
\small{$d_N$ [pc]}  &\small{1706} &\small{1350}& \small{1755}& \small{1328}& \small{1676} & \small{1872}  \\  
\small{}  &\small{+33} &\small{+45}& \small{+23}& \small{+42}& \small{+34} & \small{+36}  \\
\small{}  &\small{-32} &\small{-59}& \small{-19}& \small{-42}& \small{-39} & \small{-40}  \\
\hline \\[-1.5ex]
\small{$\sigma_N$ [pc]}  &\small{268} &\small{33}& \small{31}& \small{32}& \small{34} & \small{24}  \\
\small{}  &\small{+41} &\small{+23}& \small{+26}& \small{+18}& \small{+13} & \small{+11}  \\
\small{}  &\small{-39} &\small{-16}& \small{-17}& \small{-16}& \small{-13} & \small{-11}  \\
\hline \\[-1.5ex]
\small{Fraction [\%]}  &\small{100} &\small{19}& \small{81}& \small{11}& \small{50} & \small{39}  \\
\hline
\end{tabular}
\end{table}

The posterior distribution was sampled using the Markov Chain Monte Carlo (MCMC) affine-invariant ensemble sampler {\it emcee} \citep{emcee} with 500 walkers and 10,000 iterations. The model was compared to the observations using an unbinned maximum likelihood test. The posterior distributions were found to follow a normal distribution, and thus the median value of each parameter was used as the best fit, with the 16$^{th}$ and 84$^{th}$ percentiles used for the 1$\sigma$ uncertainties.

\section{Results}\label{sec:results}

 We applied the Shapiro-Wilk test \citep{shapiro1965} to the observed parallax distribution, which evidences that it does not follow a single normal distribution. The p-value returned ($10^{-27}$) rejects the null hypothesis that the data come from a single normally distributed population.
We then fit the observed distribution with both 2- and 3- component models (see Fig.~\ref{fig:models} and Table~\ref{table1}) and determine which model provides the best fit using the Bayesian information criterion \citep[BIC, see][]{schwarz1978}, which applies a penalty to the likelihood of more complex models so that models with different numbers of parameters can be compared.

We find that the 2-component model provides the lowest BIC and, therefore, the best fit to the data. Fig.~\ref{fig:models} corroborates that the observed parallax distribution does not fit well with a single component, and the 3-component one does not offer enough improvement. Hence we do not investigate more complex models and choose the 2-component model as representative of the observed distribution. Two different groups can be clearly distinguished, with approximate central distances of $1350\substack{+45 \\ -60}$ (rand) $^{+210}_{-160}$ (syst.) pc and  $1755\substack{+23 \\ -19}$ (rand) $^{+373}_{-261}$ (syst.) pc (systematic uncertainties take into account the 0.1~mas systematic parallax uncertainty in {\it Gaia} DR2), showing a significant distance separation between the two groups.

\begin{figure}
\centering
    \includegraphics[width=7cm,height=4.5cm]{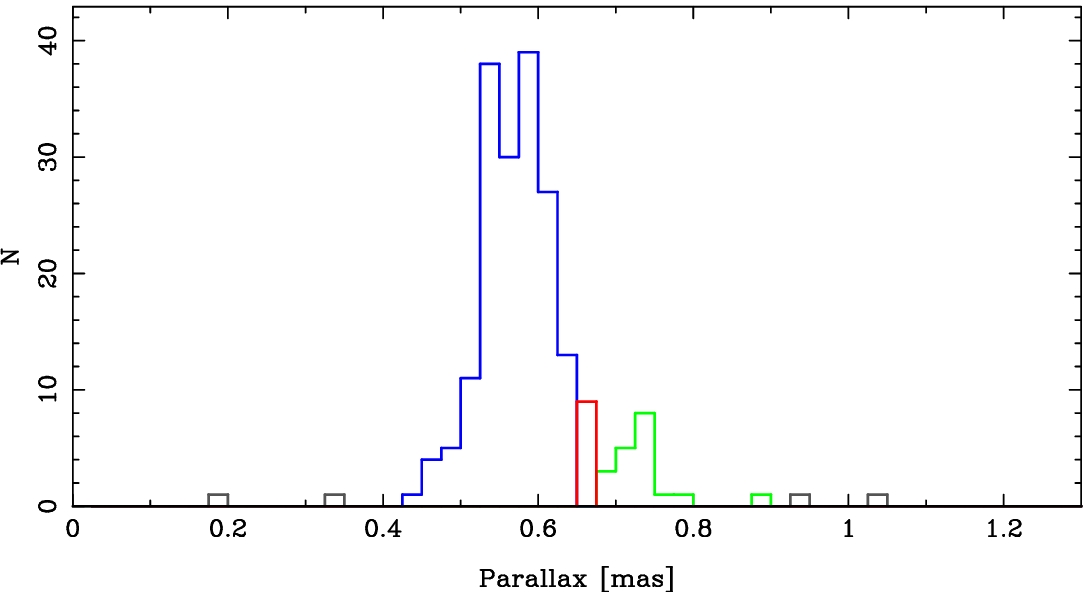}
    \caption{ Stellar sample subdivided and colour-coded by membership group. Groups 1 and 2 are represented with green and blue colour, respectively, while red indicates sources with parallaxes between those of Groups 1 and 2 that can not be placed in confidently assigned to either group (Group 0). Grey represents foreground and background contaminants (Group 3).}
    \label{fig:membership}
\end{figure}

Based on our 2-component model fit we calculated, for each star, membership probabilities for each of the populations: the foreground group (at $\sim$1350~pc, henceforth Group 1), the main group (at $\sim$1760~pc, henceforth Group 2), and whether they are foreground or background contaminants (Group 3). We then assign stars to each of these classes based upon their membership probabilities. If a star has a $>$ 75\% probability of belonging to group 1 or 2 then it is assigned to that group. For a star to be flagged as a foreground or background contaminant we require a higher probability (or effectively a lower probability that it is not a member of the other groups) of $>$ 99\%. And finally, there is a group of objects which we can not reliably place in any group (Group 0).
Figure~\ref{fig:membership} shows the parallax distribution of the sources in each group, coloured green (Group 1), blue (Group 2), grey (Group 3) or red (Group 0).
Membership groups of the final stellar sample are available in electronic form at the CDS and at MNRASL online.

While {\it Gaia} DR2 data is not as well characterised in the Galactic Plane as out of it, for the observed substructure to originate from errors or biases in the data would require systematic offsets of at least 0.2~mas in parallax, significantly larger than any quoted uncertainties or systematics in the data \citep{brown18}. We can also find no difference in the distributions of RUWE values or parallax uncertainties between the stars in the two main groups.

\section{Discussion}\label{sec:discussion}

\subsection{Spatial structure}

  We have modelled the parallax distribution of Cygnus OB2, resolving for the first time its spatial structure along the line-of-sight. Although our analysis is restricted to the OB population, \cite{wright14} showed that low- and high-mass stars are distributed in the same way, without evidence of mass segregation. We have distinguished between two clusterings, distributed on the sky as shown in Fig.~\ref{fig:sample}.
The centres of the two groups projected on the sky are not very different. Given the low density and extended nature of the foreground population it is possible that it extends beyond our field of view.  The statistical parameters obtained for each group distribution ($d_N$ and $\sigma_N$ of Model 2, see Table~\ref{table1}) suggest that the two groups are spatially separate. We consider the larger population to be the main Cygnus~OB2 association (Group 2) and consider the foreground population to be a separate group approximately $\sim$400~pc in the foreground (Group 1). 

The distance of the foreground group of $\sim$1350 pc puts it at a similar distance to Cygnus-X as a whole  \citep[see][]{rygl12} suggesting that the main part of Cygnus OB2 is actually behind Cygnus-X by several hundred parsecs (though the line-of-sight depth of Cygnus-X is not well constrained). Consequently the main group is more distant than previously thought, and therefore its stellar content will both be more luminous (approximately 1.5 times more luminous compared to the estimates in \cite{wright15}) and more massive. Interestingly this puts the distance to the main part of Cygnus~OB2 closer to that originally derived by \citet{mt91}.

\subsection{The foreground group}

We have identified 19 stars in the foreground group ($\sim$10$\%$ of the sample), seven of them classified as O-type stars. The bright BD+40 4212 double system ($G = 9.39$ mag) is included in this group, as well as the star HD 195213 ($G = 8.38$ mag). This group includes approximately 10$\%$ of the total population of O-type stars in Cygnus~OB2 and thus its total mass can be estimated as a similar fraction of the total mass of 16500~M$_\odot$ estimated by \citet{wright15}, i.e., 1650~M$_\odot$, similar to that of the Orion Nebula Cluster (although according to our results, the estimation by \cite{wright15} will have to be corrected upwards). We note that the foreground group is appreciably more dispersed on the sky than the main group. The proper motions also suggest it to be more diffuse and less likely to be a bound group. This could suggest that it is part of older foreground population that extends further outside our field of view. However, a detailed study of the physical properties of its stellar content is needed to establish the most probable scenario.

\subsection{Potential contaminants}

Here we discuss the sources identified as probable foreground or background contaminants (Group 3) and not part of either the main Cyg~OB2 population or the foreground group.

\begin{itemize}

\item{{\it Foreground contaminants}: HD 196305 is a very luminous star and has a parallax that places it at a distance of $333^{+5}_{-5}$~pc, in agreement with previous studies that suggest it to be a foreground contaminant \citep{chen13}. CCDMJ20323+4152AB has been reported as a visual double star by \cite{gili01} and, therefore, its binary nature could be affecting the parallax. MT91-426 and MT-170 also appear as foreground sources, despite the fact that \cite{wright15} proposed them as background sources based on their position in the Hertzsprung-Russell diagram. This could suggest either erroneous photometry of spectral classification, particularly in the luminosity class (e.g., a subdwarf nature).}

\item{{\it Background contaminants}:
J20272428+4115458 was classified as a B0IV star by \cite{berlanas18b} for which {\it Gaia} DR2 provides a parallax value of 0.35 $\pm$ 0.03 mas. It has a G magnitude of 11.4 mag, so the parallax uncertainty could be underestimated by up to 30$\%$. If we also add in possible systematic errors, this star is compatible with the main Cygnus~OB2 population but tentatively we suggest it as a background contaminant. 
For MT91-459 (J20331433+4119331) {\it Gaia} DR2 provides a parallax of 0.19 $\pm$ 0.04 mas, clearly indicating a background contaminant.}  

\end{itemize}

Although the highly massive, reddened and luminous Cyg OB2 \#12 hypergiant has been discarded by the astrometric selection criteria (RUWE $=$ 1.56 for this star) we highlight that {\it Gaia} DR2 places it significantly in the foreground at a distance of $840^{+105}_{-85}$~pc \citep{bailer18}. 
There are good reasons to doubt such a small inferred distance: the star has a peculiar spectrum suggesting very high luminosity and a large extinction \citep[e.g.][]{clark12}; the astrometry could reflect light centre variations in what is potentially a large angular-diameter object \citep[see][]{salas15}. Given these issues it is appropriate that it has been excluded here.

\section{Conclusions}\label{sec:conclusions}

The structure of  young  star  clusters  and  associations  is  fundamental  to  our  understanding  of  their formation and  dynamical  evolution, as well as of their stellar  content. In this work we have used {\it Gaia} DR2 parallaxes to study the 3-dimensional structure of the Cygnus~OB2 association, finding significant spatial substructure along the line-of-sight.

We fitted the observed parallax distribution with both 1-, 2- and 3- component Gaussian models and find that the best fit to the data was provided by the 2-component model, obtaining median distances to the two components of $1350\substack{+45 \\ -60}$ (rand) $^{+210}_{-160}$ (syst.) pc and  $1755\substack{+23 \\ -19}$ (rand) $^{+373}_{-261}$ (syst.) pc. The main Cygnus~OB2 group appears to be at a greater distance than has recently been thought (implying its stellar content is therefore brighter and more massive). Furthermore the parallax distribution observed suggests there may be further substructure within the association, though this is not well resolved by the available parallaxes.
The foreground group, constituting approximately 10\% of the stellar content, is several hundred parsecs in the foreground and appears more extended than the main group. A further six stars have also been found as possible background or foreground contaminants, unrelated to either group.

{\it Gaia} DR2 has provided a new view of the Cygnus OB2 association. The distance spread and substructure found within the association have shown previous concerns over the line-of-sight extent of the region were warranted. The better vision we now have moves us closer to a complete understanding of the origin and evolution of Cygnus OB2, Cygnus-X and OB associations.

\section*{Acknowledgements}
We thank J.H.J. de Bruijne, X. Luri and J. Ma\'iz-Apellaniz for helpful discussions and comments that helped improve this work. SRB and AHD acknowledge financial support from the Spanish Ministry of Science, Innovation and Universities (MCIU) under the grants AYA2015-68012-C2-01 and SEV-2015-0548, and the Gobierno de Canarias under the grant ProID-2017010115. NJW acknowledges an STFC Ernest Rutherford Fellowship (grant number ST/M005569/1). JED's research is supported via STFC grant ST/M001008/1.

%%%%%%%%%%%%%%%%%%%%%%%%%%%%%%%%%%%%%%%%%%%%%%%%%%



%\bibliographystyle{mnras}


%%%%%%%%%%%%%%%%%%%%%%%%%%%%%%%%%%%%%%%%%%%%%%%%%%

%%%%%%%%%%%%%%%%% APPENDICES %%%%%%%%%%%%%%%%%%%%%

\section*{Supporting information}\label{sect:online}
The list of Cygnus~OB2 sources used for this work, derived membership and those stars that have not passed the selection criteria are available in electronic form at the CDS and at MNRAS online.

%%%%%%%%%%%%%%%%%%%%%%%%%%%%%%%%%%%%%%%%%%%%%%%%%%

% Don't change these lines
\bsp	% typesetting comment
\label{lastpage}
\end{document}